\documentclass[11pt,twoside]{article}

\usepackage{asp2006}
\usepackage{epsf}
\usepackage{psfig}
\usepackage{lscape}
\usepackage{graphics}

\begin{document}
\title{The Impact of Stellar Collisions in the Galactic Center }   
\author{M.B. Davies, R.P. Church, D. Malmberg, and S. Nzoke}   
\affil{Lund Observatory, Box 43, SE-221 00, Lund, Sweden}    
\author{J. Dale}   
\affil{Astronomical Institute, Academy of Sciences of the Czech Republic, Bocni II 1401/2a, 141 31 Praha 4}    
\author{M. Freitag}   
\affil{Gymnase de Nyon, 
Route de Divonne 8, 1260 Nyon, Switzerland}    

\begin{abstract} 
We consider whether stellar collisions can explain the observed depletion
of red giants in the Galactic center. We model the stellar population with two
different IMFs: 1) the Miller-Scalo and 2) a much flatter IMF. In the former case,
low-mass main-sequence stars dominate the population, and collisions are unable
to remove red giants out to 0.4\,pc although brighter red giants much closer in
may be depleted via collisions with stellar-mass black holes. For a much flatter IMF,
the stellar population is dominated by compact remnants ({\it i.e.}\ black holes, white dwarfs
and neutron stars). The most common collisions are then those between 
main-sequence stars and compact remnants. Such encounters are likely to destroy
the main-sequence stars and thus prevent their evolution into red giants. In this
way, the red-giant population could be depleted out to 0.4\,pc matching observations.
If this is the case, it implies the Galactic center contains a much larger population
of stellar-mass black holes than would be expected from a regular IMF. This may
in turn have implications for the formation and growth of the central supermassive
black hole.
\end{abstract}

\section{Introduction}

The Milky Way  contains a supermassive black hole at its very center,
Sagittarius A*, whose mass is $\approx4\times10^{6}$\,M$_{\odot}$ \citep[e.g.][]{mdavies_schodel03}.  
A dense stellar cluster surrounds this black hole, with a central density at least comparable
to that seen in the cores of the densest globular clusters. At least one, and possibly two, discs
containing young stars at distances between $\sim0.04$ and $\sim0.3$\,pc from the black hole
are also seen \citep{mdavies_paumard06}. The latter stellar population is thought to have an unusually
flat IMF \citep{mdavies_paumard06}.  It has long been known that the central 0.2\,pc or so of the Galactic 
center is deficient in bright red giants \citep{mdavies_genzel96}.  Since the center of the Galaxy 
has a high number density of stars, it is natural to suggest that stellar collisions may explain
the observed depletion.

A recent crop of papers have studied the stellar population in the central
regions \citep{mdavies_bartko10, mdavies_buchholz09, mdavies_do09}.  They report
that the early-type stars (bright main-sequence stars) follow a cusp-like
profile whilst the surface density of late-type (red-giant) stars that are
brighter than a K magnitude of 15.5 is rather flat out to 0.4\,pc or 10\,arcsec.
This surprising result implies that the red-giant population is depleted out to
about 0.4\,pc from the central supermassive black hole.  We will consider here
whether stellar collisions could be responsible for this observed depletion.  We
will consider two different cases: 1) where the stellar population in the
Galactic center is drawn from a Miller-Scalo IMF, and 2) where the IMF is much
flatter. In each case, we calculate collision probabilities between the various
stellar species, and produce, via Monte Carlo techniques, a stellar population.
From this we measure the surface density of the early and late-type stars (to
compare directly with what is observed) making some reasonable assumptions
concerning the effects of collisions on the population.

\begin{figure}[!ht]
\plotfiddle{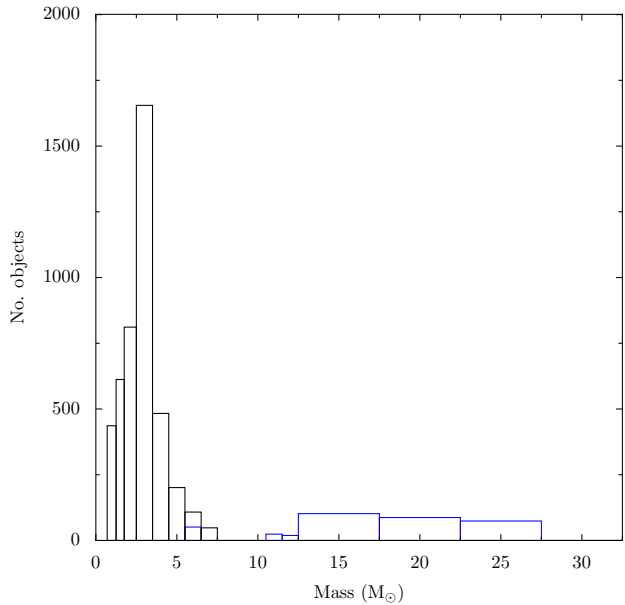}{8truecm}{0.0}{70}{70}{-100}{20}
\caption{A plot of the distribution of stellar masses contributing to the
observed early-type (between 12 and 27\,M$_{\odot}$)  and late-type stars
(between 1 and 5\,M$_{\odot}$) at the Galactic center, assuming the stellar
population is drawn from a flat IMF ($\Gamma= 1.0$).}
\label{mdavies_figure1}
\end{figure}

The masses of stars contributing to the observed early and late-type populations
are rather different, as shown in Fig.~\ref{mdavies_figure1}, where we have
produced a synthetic population with a flat IMF. We see that the early-type
stars are virtually all between 12 and 27\,M$_{\odot}$ whereas stars
contributing to the observed late-type population have much lower masses,
between 1 and 5\,M$_{\odot}$.

\section{The Effects of Crowdedness}

The Galactic center is a crowded environment, with number densities of stars probably
of order $10^6$\,stars\,pc$^{-3}$ or more. In such dense environments, collisions between
stars will be frequent, and thus affect the stellar population. 
The cross section for two stars, having a relative velocity at infinity
of $V_\infty$, to pass within a distance $R_{\rm min}$ is given by

\begin{equation}
\sigma = \pi R_{\rm min}^2 \left( 1 + {V^2 \over V_\infty^2} \right)
\end{equation}

\noindent where $V$ is the relative velocity of the two
stars at closest approach in a parabolic encounter ({\it i.e.}\ $V^2 = 2 G (M_1 + M_2)/R_{\rm min}$, where $M_1$ and $M_2$ are the masses
of the two stars).
The second term is due to the attractive gravitational force between the 
two stars, and
is referred to as the gravitational focusing term. In the very center of the Galaxy, where
the supermassive black hole dominates, and stars may be assumed to
move on Keplerian orbits around the black hole at speeds exceeding 
1000\,km/s,  $V \ll V_\infty$ and we recover
the result, $\sigma \propto R_{\rm min}^2$. In this regime collisions involving
larger red giants will be relatively more frequent despite their short
lifetimes compared to main-sequence stars. This is not the case in globular clusters where 
 $V \gg V_\infty$ and thus  $\sigma \propto R_{\rm min}$.
One may estimate the timescale for a given star to undergo an encounter
with another star,
$\tau_{\rm coll} = 1/n \sigma v$. The collision timescale will therefore be a
function of both the number density of stars and the makeup of the stellar
population, {\it i.e.}\ the distribution of stellar masses and types.

The effects of collisions will depend on the types of stars involved  and on the
relative speed of the two stars when they collide.  Collisions involving two
main-sequence stars occurring at relatively low speed (less than the surface
escape speed of the stars) are likely to result in the merger of the two objects
with relatively low amounts of mass loss. Collisions occurring at much higher
speeds are likely to lead to significant mass loss and even to the destruction
of the stars involved. 

\begin{figure}[!ht]
\plotone{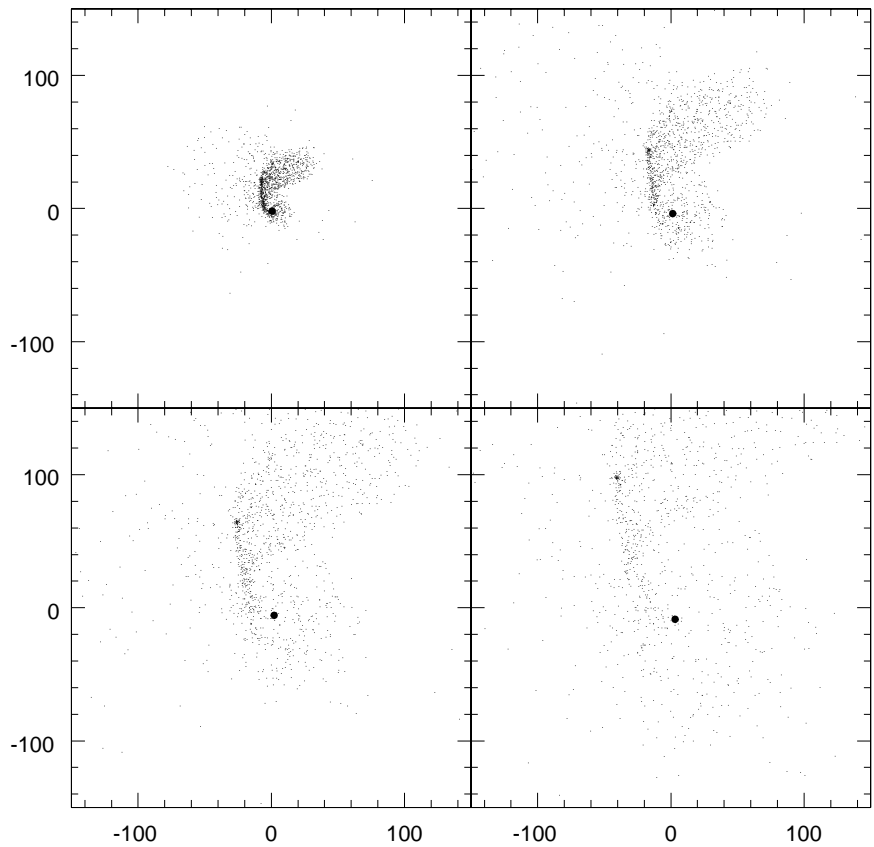}
\caption{A series of snap-shots of a collision between a main-sequence star and a stellar-mass
black hole using our SPH code. The position of the black hole is given by the black filled circle
and we show only  1\% of the SPH particles.}
\label{mdavies_figure2}
\end{figure}

Collisions between main-sequence stars and compact objects ({\it i.e.}\ black
holes, white dwarfs or neutron stars) are likely to be destructive (Dale et al.,
in preparation).  In the case of a black hole impactor, the main-sequence star
is often torn apart by tidal forces, with some material being accreted by the
black hole and rest being ejected.  This is illustrated in
Fig.~\ref{mdavies_figure2} where we show snapshots of a collision between a
1\,M$_\odot$ main-sequence star and a 10\,M$_\odot$ black hole. The close
passage of the black hole results in the tidal disruption of the main-sequence
star. A small fraction of the material is accreted by the black hole and the
rest is dispersed.  For neutron star and white dwarf impactors, a larger
fraction of the material from the main-sequence star may form an envelope around
the compact object. However such an object is likely to be relatively short
lived, perhaps appearing as a bright red supergiant, before the envelope is
ejected. Thus all collisions between main-sequence stars and compact objects
will act to reduce the population of luminous stars.

\begin{figure}[!ht]
\plotone{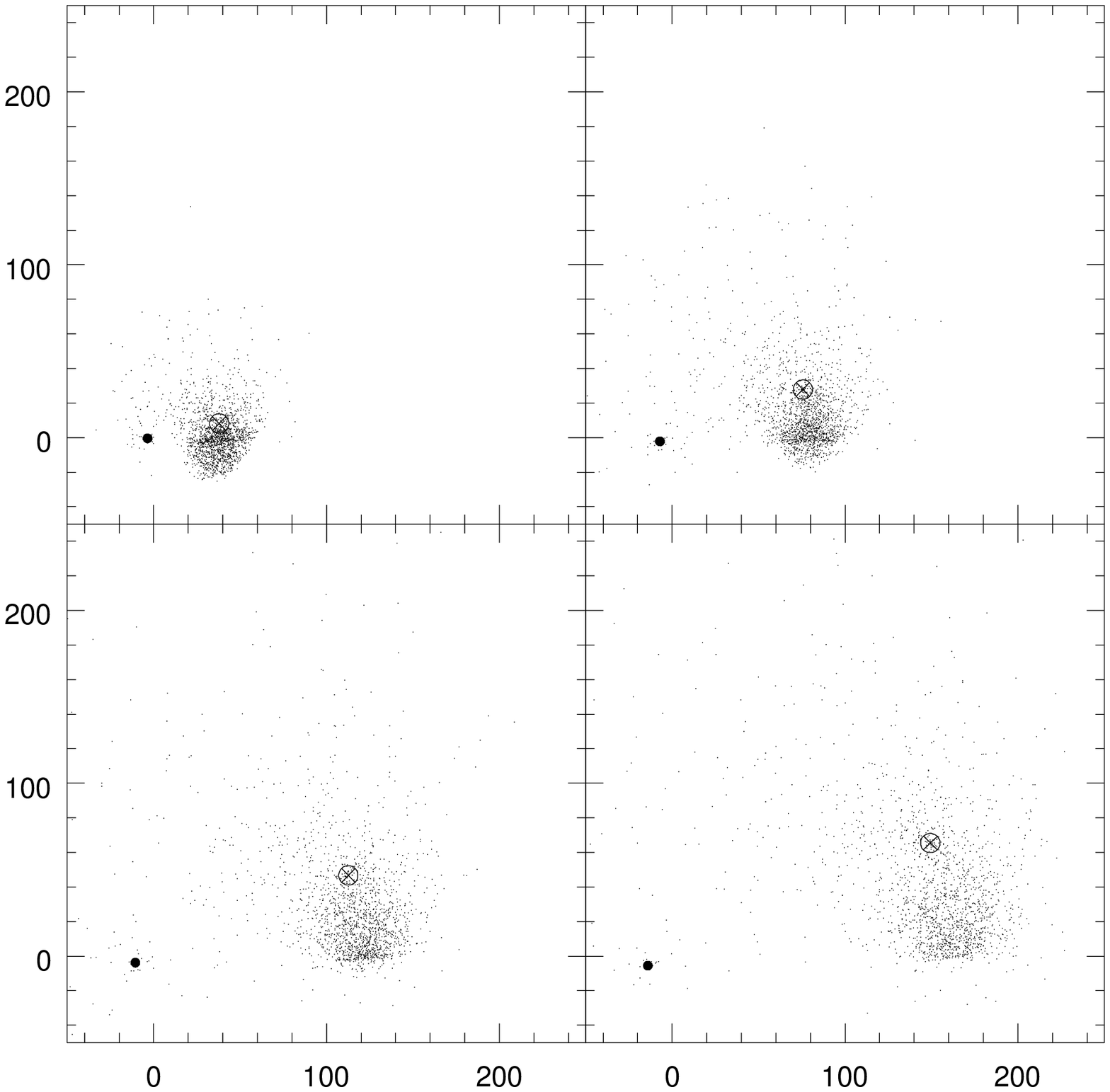}
\caption{A series of snap-shots of a collision between a red giant and a stellar-mass
black hole using our SPH code. The black hole is labelled with a black filled circle
and the red-giant core is labelled with an X in the center of an open circle. For clarity in the
plot, only 1\% of the SPH particles are shown.}
\label{mdavies_figure3}
\end{figure}

Even though they are very frequent, collisions
between red giants and main-sequence stars in fact have very little effect, as the main-sequence
star passes through the envelope and effects very little mass loss \citep{mdavies_bailey99}.
 The same is also true
for white dwarf or neutron star impactors.
 More interesting are encounters involving black holes.
In Fig.~\ref{mdavies_figure3} we show snapshots of an encounter between a 1\,M$_\odot$ red giant and 
and 10\,M$_\odot$ black hole ($V_\infty =  800$ km/s, $r_{\rm min} = 10$ R$_\odot$). As the black hole
passes close to the red-giant core, it gives it a jerk and the core is ejected at high speed, 
retaining only a small fraction of the envelope (in this case about 13\%). Such excessive mass
loss will prevent this red giant from becoming brighter: we have thus removed it from the pool
of brighter red giants.

Thus we see that in order to deplete the stellar population of red giants (as indicated
by the observations), we must consider collisions of three types: 1) encounters between
red giants and black holes, 2) encounters between two main-sequence stars, and 3)
encounters between main-sequence stars and compact objects
({\it i.e.}\ black holes, white dwarfs and neutron stars).
The relative frequencies of encounters will depend on the IMF of the underlying stellar
population. With a Miller-Scalo type IMF, where most stars are of low mass, the majority
of stars formed will still be on the main sequence today, with relatively few stars
having evolved off the main-sequence to ultimately form compact remnants.
 Thus collisions involving two main-sequence
stars will be more frequent than collisions between main-sequence stars and compact
objects. Collisions between red giants and black holes may be relatively frequent, at least
in the very center, owing to the larger size (and thus collisional cross section) of red giants.

\section{Collisions for a Miller-Scalo IMF}

\begin{figure}[!ht]
\plotfiddle{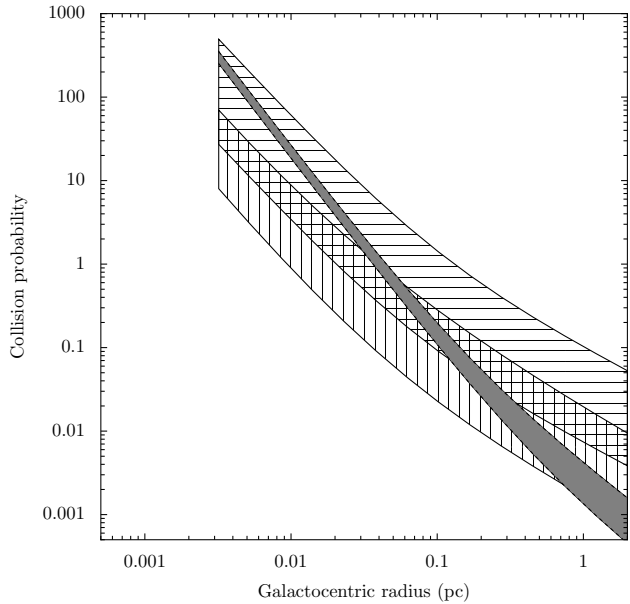}{8.4truecm}{0.0}{70}{70}{-100}{20}
\caption{The collision rates between the various stellar species. 
Here  the collision probability is the expected collision rate integrated 
over the entire lifetime of the main-sequence or red-giant phase. In other
words, the star has a good chance of being involved in a collision if the collision
probability is one.
The gray band
is for encounters between red giants and black holes,  the vertically-shaded region
is for encounters between main-sequence stars and compact objects, 
and the horizontally-shaded region is for encounters between two main-sequence stars.
In all cases, we consider stars between 1.0\,M$_\odot$ and 5.0\,M$_\odot$.
The stellar population has a Miller-Scalo IMF.}
\label{mdavies_figure4}
\end{figure}

We consider first that the stellar population in the Galactic center has been drawn from
a Miller-Scalo IMF \citep{mdavies_miller79} where we also assume
the stars have a uniform spread in ages from 0 to 14\,Gyr. Using a density distribution for the stars given 
by $n_{\rm *}(r)\propto r^{-1.4}$  and a density distribution for 10\,M$_\odot$ 
black holes of $n_{\rm BH}\propto r^{-1.8}$ \citep{mdavies_freitag06}, we compute the collision probabilities between the various
stellar types as  a function of Galactocentric radius. The results are shown in Fig.~\ref{mdavies_figure4}. We see that collisions
between red giants and  black holes are the most frequent in the very central regions
whilst  collisions between two main-sequence stars dominate further out. 
For stars following a Miller-Scalo IMF, 
collisions between main-sequence stars and black holes (or other compact remnants) are
relatively less frequent and are unlikely  to have a significant effect on the stellar population
apart from perhaps in the very central regions. 
Even though collisions between 
main-sequence stars are relatively frequent, in the case of a Miller-Scalo IMF
with a low-mass cutoff of 0.2\,M$_\odot$ the number of main-sequence stars in the mass
range 1--4\,M$_\odot$ will not be reduced. The reason is that although some stars in this
mass range will be removed as they merge with other stars to produce more massive
stars, they will be replaced by stars produced by the merger of lower-mass stars. This process will be investigated in detail in Dale et al.~(in preparation). It could be that a stellar population
with a Miller-Scalo IMF but with a larger low-mass cutoff may lead to a depletion of the lower-mass
main-sequence stars which later evolve to produce the observed population of red giants.

Collisions between red giants and stellar-mass black holes
are able to deplete the red-giant population within
the inner 0.08\,pc or so but not further out.  Collisions can plausibly explain the depletion of
red giants of middle brightness ($10.5 < K < 12$) but not those in the brightest band
($K > 10.5$) which are seen to be depleted to about 0.2\,pc \citep{mdavies_genzel03, mdavies_dale09}.
Collisions are not able to explain the observed flattening of the red-giant population
seen out to 0.4\,pc \citep{mdavies_bartko10,mdavies_buchholz09}.

\section{Collisions for a Flat IMF}

In this section we consider the case where the stellar population in the Galactic center
is drawn from a much flatter IMF than a Miller-Scalo IMF.  Observations of the young disc
of stars in the Galactic center suggest that they may have been drawn from such
an IMF with a power-low slope as low as $\Gamma = 0.45$ \citep{mdavies_bartko10}.  A stellar population
drawn from such a flat IMF would be quite different from one drawn from a Miller-Scalo IMF.
A much larger fraction of stars would be massive, and explode as core-collapse supernovae
producing either black holes or neutron stars. In addition, assuming stars have an approximately
uniform spread in ages, a very large fraction of all stars will have evolved to become compact
remnants. ({\it i.e.}~black holes, white dwarfs, or neutron stars). In other words, the majority of stellar
objects will in fact be compact objects. The most common flavor of collision will then be 
collisions involving compact objects.

\begin{figure}[!ht]
\plotfiddle{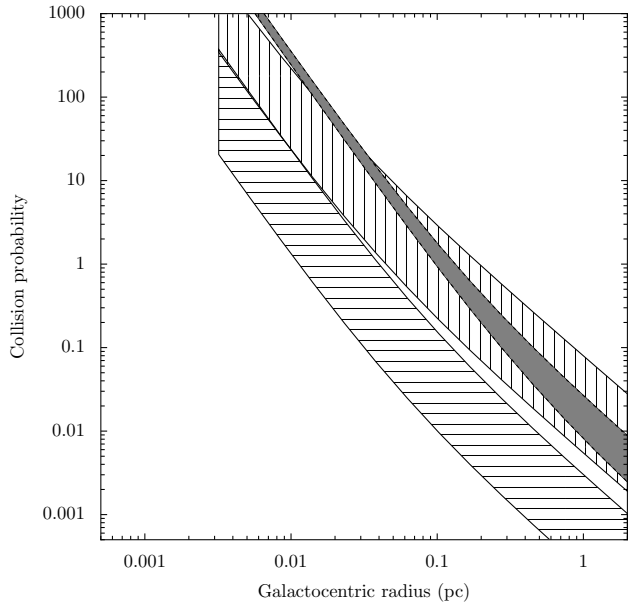}{8.4truecm}{0.0}{70}{70}{-100}{20}
\caption{The collision rates between the various stellar species. 
Here  the collision probability is the expected collision rate integrated 
over the entire lifetime of the main-sequence or red-giant phase. In other
words, the star has a good chance of being involved in a collision if the collision
probability is one.
The gray band
is for encounters between red giants and black holes,  the vertically-shaded region
is for encounters between main-sequence stars and compact objects, 
and the horizontally-shaded region is for encounters between two main-sequence stars.
In all cases, we consider stars between 1.0\,M$_\odot$ and 5.0\,M$_\odot$.
The stellar population has a flat IMF ($\Gamma = 1.0$).}
\label{mdavies_figure5}
\end{figure}

In order to produce a stellar population, we assume a model where a stellar disc,  similar
to the one seen today, is produced about every $3 \times 10^{7}$ years up to 14\,Gyr ago, with the
most recent disc being produced only 6 million years ago to be consistent
with observations.  
The disc mass is set to match the observations of the early-type stars which come entirely
from the most-recent disc, whilst the total number of discs produced is set to match
the observed late-type population 10\,arcsec from the supermassive black hole. One should
note that the formation mechanism for such discs of stars is unclear, one suggestion being that
gas clouds interact tidally with the supermassive black hole \citep{mdavies_bonnell08}.
We consider here for purposes
of illustration  the case of  $\Gamma = 1.0$. For this IMF (which is flat in log mass), 
the mass is each disc is
$10^{4}$\,M$_\odot$, with the stars having a range of masses between 1.0 and 32.5\,M$_\odot$.
 It is entirely possible that the star formation rate was higher in the past.
As a simple first model, we assume that in addition to the discs produced at a steady rate,
there was an excess of discs produced sufficiently early in the history of the Galaxy that essentially
all the stars from these earlier discs have evolved into compact remnants. The exact excess
of compact objects will depend on the history of the disc formation rate, 
but here we assume that two thirds of all star formation occurred at this earlier epoch.
This will increase the number of compact objects present in the Galactic center population.
As collisions involving compact remnants destroy main-sequence stars,
increasing the compact object population will enhance the red-giant depletion. 

The collision probability for collisions between stars of various types for the stellar
population produced as described above are shown in Fig.~\ref{mdavies_figure5}.
 We see here that collisions between main-sequence
stars and  compact objects are most likely (at least further out from the center). 
Main-sequence stars are likely to undergo such a collision, and thus be destroyed,
out to a distance of about 0.2\,pc.

As discussed earlier, collisions between compact objects
and main-sequence stars are likely to be destructive. 
Therefore with a flat IMF, and a suitably
dense stellar population, the low-mass main-sequence
stars which would ordinarily have evolved to become
the observed late-type population are depleted within a distance of about 0.2\,pc of the supermassive 
black hole. Therefore a  stellar population produced from a flat IMF may be able to explain
the flat surface density profile which has been observed for the late-type stars (red giants) as 
stars in the mass range 1-5\,M$_\odot$ within 0.2\,pc have simply been destroyed before
they could evolve into red giants.

\section{Discussion}

\begin{figure}[!ht]
\plottwo{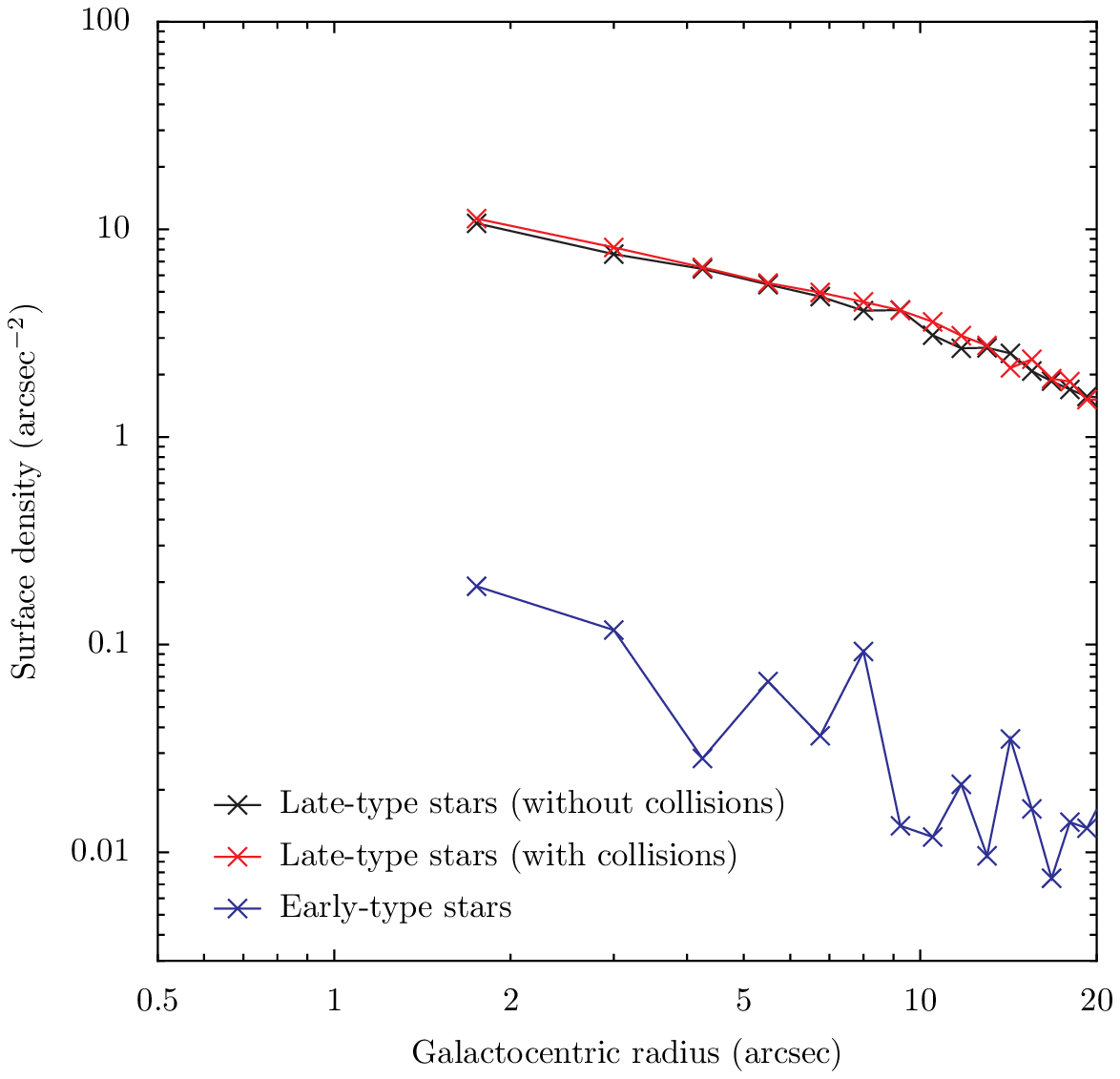}{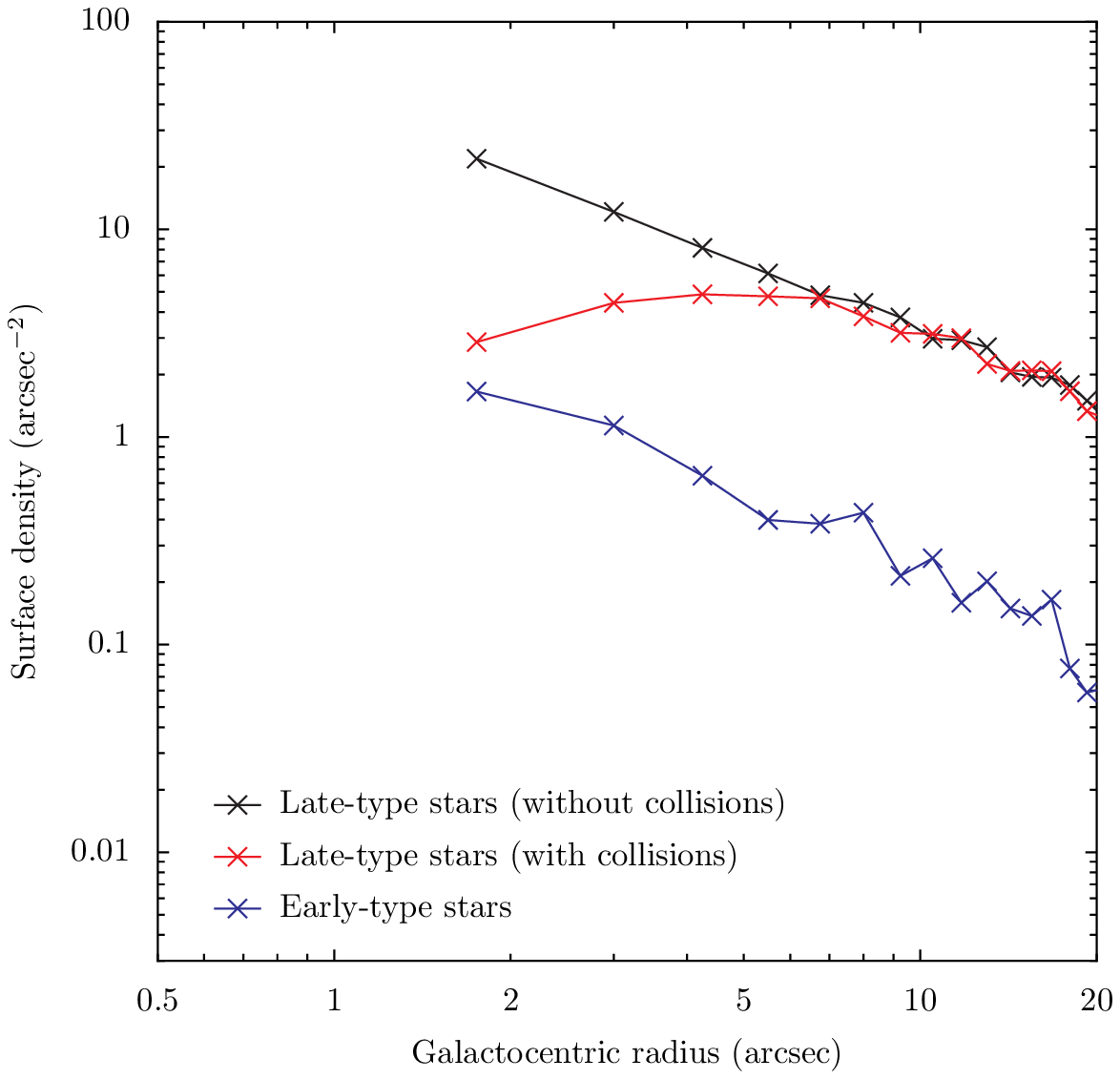}
\caption{The surface density of early and late-type stars and the depletion due stellar collisions
for a) Miller-Scalo IMF, and b) a flat IMF ($\Gamma=1.0$). In both cases, the top line is for late-type stars (without
allowing for destructive collisions), the middle line is for late-type stars (allowing for destructive collisions), and the bottom line is for the early-type stars. Note that these plots are the results of single Monte Carlo realisations, so that each line has an associated uncertainty.}
\label{mdavies_figure6}
\end{figure}

Equipped with the collision probabilities as a function of radius, we are now able
to synthesize a stellar population for both Miller-Scalo and flat IMFs allowing for
the collisional depletion of the red-giant population. In Fig.~\ref{mdavies_figure6} we show the
calculated surface density profiles for early-type stars 
and late-type stars having K magnitudes brighter than 15.5  for both IMFs. 
In both cases, the stellar population has been normalised to the surface brightness
of late-type stars 10\,arcsec from the Galactic center,  seen to be about 3 sources per square arcsec
(this normalization was also applied to the collision probability calculations and figures
shown earlier). 
In both plots we do not include the effect of stellar collisions on the population of
early-type stars as the collisional depletion of these massive stars will be very small.
However it should be noted that the existence of the so-called S stars in the very center
of the galaxy (distances less than 1 arcsecond from the central black hole) may place
limits on the number density of compact objects.
The plots shown in  Fig.~\ref{mdavies_figure6} 
should be compared to Fig.~11 of \citet{mdavies_buchholz09}.
One can see from the plots that the early-type population only matches the observations 
in the case of the flat IMF. A Miller-Scalo population is clearly not able to consistently match
both the observed early and late populations.
It is also clear
from these plots that in the case of the Miller-Scalo IMF, the effect of stellar collisions on the red-giant population
is negligible. However for the flat IMF, appreciable red-giant depletion occurs out to 10\,arcseconds.
Thus a flat IMF would seem to have the potential to explain the observed (flat) surface
density profile of the red giants.

One problem with the model used here is that the total
stellar mass out to 0.4\,pc is about $5 \times 10^6$\,M$_\odot$. This is much higher
than suggested by observations \citep{mdavies_schodel03}.  However, in our calculations 
using a flat IMF here
we have made the unrealistic assumption that stars remain at the radius at which they 
were formed. In other words, we have not allowed for the effects of mass segregation.
In reality, the large population of
heavier stellar-mass black holes will sink in the potential,
forming a central sub-cluster.  This sub-cluster may then collapse to form a supermassive black hole in the manner
envisaged by \cite{mdavies_quinlan87} if a supermassive black hole is not there already,
or the stellar-mass black holes may simply be fed into an existing supermassive black hole.
In either case, a large fraction of the stellar-mass black holes may end up inside the supermassive
black hole. It is interesting in this context to note that the total mass contained in stellar-mass
black holes for our flat IMF (about $3.5 \times 10^6$\,M$_\odot$) is close to the observed
value for the supermassive black-hole mass. 
A larger population of black holes produced by a flat IMF would also lead 
to a much higher rate of EMRI--type events, where stellar-mass black holes spiral in to the 
central, supermassive black hole, emitting gravitational radiation in the process. EMRIs
could be an  important source for LISA.
Our calculations here have been a simplification. We have made the unrealistic
assumption that stars remain at the radius they were formed. In addition we have not
allowed for the growth of the supermassive black hole. Early on, before the 
supermassive black hole has grown, the disc-mode of star formation may be different or not
occur at all owing to a lack of a supermassive black hole. A natural next step
in these calculations will be inclusion of the dynamical evolution of the stellar
cluster  and for 
the growth of the central black hole (Nzoke et al., in preparation).



\acknowledgements RPC acknowledges support from the Wenner Gren Foundation. 
JED is supported by a Marie Curie fellowship as part of the European Commission FP6 Research Training Network `Constellation' under contract MRTN--CT--2006--035890 and the Institutional Research Plan AV0Z10030501 of the Academy of Sciences of the Czech Republic and project LC06014--Centre for Theoretical Astrophysics of the Ministry of Education, Youth and Sports of the Czech Republic.



\begin{thebibliography}{}

\bibitem[\protect\citeauthoryear{{Bailey} \& {Davies}}{{Bailey} \&
  {Davies}}{1999}]{mdavies_bailey99}
{Bailey} V.~C.,  {Davies} M.~B.,  1999, \mnras, 308, 257

 \bibitem[\protect\citeauthoryear{{Bartko}, {Martins}, {Trippe}, {Fritz},
  {Genzel}, {Ott}, {Eisenhauer}, {Gillessen}, {Paumard}, {Alexander},
  {Dodds-Eden}, {Gerhard}, {Levin}, {Mascetti} \& {Nayakshin}}{{Bartko}
  et~al.}{2010}]{mdavies_bartko10}
{Bartko} H.,  {Martins} F.,  {Trippe} S.,  {Fritz} T.~K.,  {Genzel} R.,  {Ott}
  T.,  {Eisenhauer} F.,  {Gillessen} S.,  {Paumard} T.,  {Alexander} T.,
  {Dodds-Eden} K.,  {Gerhard} O.,  {Levin} Y.,  {Mascetti} L.,    {Nayakshin}
  S.,  2010, \apj, 708, 834
  
   \bibitem[\protect\citeauthoryear{{Bonnell} \& {Rice}}{{Bonnell} \&
  {Rice}}{2008}]{mdavies_bonnell08}
{Bonnell} I.~A.,  {Rice} W.~K.~M.,  2008, \science, 321, 1060

\bibitem[\protect\citeauthoryear{{Buchholz}, {Sch{\"o}del} \&
  {Eckart}}{{Buchholz} et~al.}{2009}]{mdavies_buchholz09}
{Buchholz} R.~M.,  {Sch{\"o}del} R.,    {Eckart} A.,  2009, \aap, 499, 483

 \bibitem[\protect\citeauthoryear{{Dale}, {Davies}, {Church} \&
  {Freitag}}{{Dale} et~al.}{2009}]{mdavies_dale09}
{Dale} J.~E.,  {Davies} M.~B.,  {Church} R.~P.,    {Freitag} M.,  2009, \mnras,
  393, 1016

\bibitem[\protect\citeauthoryear{{Do}, {Ghez}, {Morris}, {Lu}, {Matthews},
  {Yelda} \& {Larkin}}{{Do} et~al.}{2009}]{mdavies_do09}
{Do} T.,  {Ghez} A.~M.,  {Morris} M.~R.,  {Lu} J.~R.,  {Matthews} K.,  {Yelda}
  S.,    {Larkin} J.,  2009, \apj, 703, 1323
  
  \bibitem[\protect\citeauthoryear{{Freitag}, {Amaro-Seoane} \&
  {Kalogera}}{{Freitag} et~al.}{2006}]{mdavies_freitag06}
{Freitag} M.,  {Amaro-Seoane} P.,    {Kalogera} V.,  2006, \apj, 649, 91

 \bibitem[\protect\citeauthoryear{{Genzel}, {Thatte}, {Krabbe}, {Kroker} \&
  {Tacconi-Garman}}{{Genzel} et~al.}{1996}]{mdavies_genzel96}
{Genzel} R.,  {Thatte} N.,  {Krabbe} A.,  {Kroker} H.,    {Tacconi-Garman}
  L.~E.,  1996, \apj, 472, 153
  
  \bibitem[\protect\citeauthoryear{{Genzel}, {Sch{\"o}del}, {Ott}, {Eisenhauer},
  {Hofmann}, {Lehnert}, {Eckart}, {Alexander}, {Sternberg}, {Lenzen},
  {Cl{\'e}net}, {Lacombe}, {Rouan}, {Renzini} \& {Tacconi-Garman}}{{Genzel}
  et~al.}{2003}]{mdavies_genzel03}
{Genzel} R.,  {Sch{\"o}del} R.,  {Ott} T.,  {Eisenhauer} F.,  {Hofmann} R.,
  {Lehnert} M.,  {Eckart} A.,  {Alexander} T.,  {Sternberg} A.,  {Lenzen} R.,
  {Cl{\'e}net} Y.,  {Lacombe} F.,  {Rouan} D.,  {Renzini} A.,
  {Tacconi-Garman} L.~E.,  2003, \apj, 594, 812

 \bibitem[\protect\citeauthoryear{{Miller} \& {Scalo}}{{Miller} \&
  {Scalo}}{1979}]{mdavies_miller79}
{Miller} G.~E.,  {Scalo} J.~M.,  1979, \apjs, 41, 513

 \bibitem[\protect\citeauthoryear{{Paumard}, {Genzel}, {Martins}, {Nayakshin},
  {Beloborodov}, {Levin}, {Trippe}, {Eisenhauer}, {Ott}, {Gillessen}, {Abuter},
  {Cuadra}, {Alexander} \& {Sternberg}}{{Paumard}
  et~al.}{2006}]{mdavies_paumard06}
{Paumard} T.,  {Genzel} R.,  {Martins} F.,  {Nayakshin} S.,  {Beloborodov}
  A.~M.,  {Levin} Y.,  {Trippe} S.,  {Eisenhauer} F.,  {Ott} T.,  {Gillessen}
  S.,  {Abuter} R.,  {Cuadra} J.,  {Alexander} T.,    {Sternberg} A.,  2006,
  \apj, 643, 1011
  
  \bibitem[\protect\citeauthoryear{{Quinlan} \& {Shapiro}}{{Quinlan} \&
  {Shapiro}}{1987}]{mdavies_quinlan87}
{Quinlan} G.~D.,  {Shapiro} S.~L.,  1987, \apj, 321, 199

\bibitem[\protect\citeauthoryear{{Sch{\"o}del}, {Ott}, {Genzel}, {Eckart},
  {Mouawad} \& {Alexander}}{{Sch{\"o}del} et~al.}{2003}]{mdavies_schodel03}
{Sch{\"o}del} R.,  {Ott} T.,  {Genzel} R.,  {Eckart} A.,  {Mouawad} N.,
  {Alexander} T.,  2003, \apj, 596, 1015
  
\end{thebibliography}
\end{document}